# The NOAO KOSMOS Data Handling System

R. Seaman
NOAO Science Data Management
8 June 2014

## 1. Introduction

KOSMOS and COSMOS[1] are twin high-efficiency imaging spectrographs that have been deployed as NOAO facility instruments for the Mayall 4-meter telescope on Kitt Peak in Arizona and for the Blanco telescope on Cerro Tololo in Chile, respectively.[2] These instruments are modified versions [Martini 2014] of the Ohio State Multi-Object Spectrograph (OSMOS) previously commissioned at the MDM Observatory. KOSMOS benefits from an all-refractive optical design that supports longslit and multi-object spectroscopy as well as imaging modes. The instrument can be rapidly reconfigured via motions of slit and disperser wheels plus dual filter wheels. Slit masks can be quickly swapped into the wheel during the night.

Efficient hardware requires efficient software to realize its full capabilities. The NOAO Data Handling System (DHS) has seen aggressive use over several years at both the Blanco and Mayall telescopes with NEWFIRM (the NOAO Extremely Wide-Field Infrared Imager[3]) and the Mosaic-1.1 wide-field optical imager[4]. Both of these instruments also rely on the Monsoon[5] array controller and related software, and on instrument-specific versions of the NOAO Observation Control System (NOCS). NOCS, Monsoon and DHS are thus a well-tested software suite that was adopted by the KOSMOS project.

This document will describe the specifics of the KOSMOS implementation of DHS, in particular in support of the original two-amplifier e2v 2Kx4K CCD detectors with which the instruments were commissioned. The emphasis will be on the general layout of the DHS software components and the flow of data and metadata through the system as received from Monsoon and the NOCS. Instructions will be provided for retrieving and building the software, and for taking simulated and actual exposures.

## 2. Observing and Software Environment

Software dependencies for KOSMOS DHS are somewhat extensive. Most fundamentally, note that the suite of NOCS, Monsoon and DHS software currently requires a 32-bit version of Linux. The KOSMOS computer rack holds four computers each running the same 32-bit OS, these are generally divided between these three classes of software, plus one common spare that can fill in for any of these roles. The observer shares the resources of the NOCS computer. During commissioning the two computer racks (for KOSMOS and COSMOS) were located in Tucson in the lab and later in the Flex Rig space. The DNS names of the various hosts and their corresponding IP addresses are shown in table 1 for the Mayall Telescope on Kitt Peak (KOSMOS) and the Blanco Telescope on Cerro Tololo (COSMOS). Since it is unlikely that COSMOS will ever return to Tucson, the Tucson IP numbers are only provided for KOSMOS.

---

[1] http://www.noao.edu/nstc/kosmos/

[2] Throughout the remainder of this document, "KOSMOS" will be taken to refer to either or both instruments unless a particular instrument is specified.

[3] http://www.noao.edu/ets/newfirm/

[4] http://www.noao.edu/kpno/mosaic/

[5] http://www.noao.edu/nstc/monsoon/

**Table 1 – Host names and IP addresses for the computer racks for each instrument. The COSMOS rack also had Tucson IP addresses during commissioning in the lab and Flex Rig, but does not appear to have assigned addresses in La Serena. The spare computers would be reassigned names and IPs as needed to fill one or another role.**

| Host | DNS | IP | Role |
|---|---|---|---|
| kosmos | kosmos-4m.kpno.noao.edu | 140.252.52.61 | Observer / NOCS |
| kosmospan | kosmospan-4m.kpno.noao.edu | 140.252.52.62 | Monsoon PAN |
| kosmosdhs | kosmosdhs-4m.kpno.noao.edu | 140.252.52.63 | DHS |
| kosmosspare | kosmosspare-4m.kpno.noao.edu | 140.252.52.64 | Spare |
| | | | |
| kosmos | kosmos-dtn.tuc.noao.edu | 140.252.22.175 | Observer / NOCS |
| kosmospan | kosmospan-dtn.tuc.noao.edu | 140.252.22.176 | Monsoon PAN |
| kosmosdhs | kosmosdhs-dtn.tuc.noao.edu | 140.252.22.177 | DHS |
| kosmosspare | kosmosspare-dtn.tuc.noao.edu | 140.252.22.178 | Spare |
| | | | |
| cosmos | cosmos-4m.ctio.noao.edu | 139.229.14.196 | Observer / NOCS |
| cosmospan | cosmospan-4m.ctio.noao.edu | 139.229.14.197 | Monsoon PAN |
| cosmosdhs | cosmosdhs-4m.ctio.noao.edu | 139.229.14.198 | DHS |
| cosmosspare | cosmosspare-4m.ctio.noao.edu | 139.229.14.199 | Spare |

A variety of Unix user accounts are used on these machines for different purposes. DHS software development shares the "monsoon" account with Monsoon activities. NOCS software development is done under the `kosmos` account (both for KOSMOS and COSMOS). Source code, documentation, and various configuration files for Monsoon and NOCS are located under the corresponding home directories of these accounts. The DHS software is installed under `/dhs`, which is typically a symbolic link to some convenient location in the file system. Various startup and runtime assumptions are made in DHS to the `/dhs` location. The observer logs in as `observer`, and generally only to the main computer (`kosmos-4m` or `cosmos-4m`). Password information may change, of course, and the usual arrangements may be made to permit ssh authentication between various accounts on various machines.

Some understanding of the instrument configuration and observing environment will also help to place KOSMOS DHS into the proper context. KOSMOS and COSMOS are mounted at the cassegrain focus of the corresponding 4-meter telescopes. Light passes directly through the instrument to the CCD dewar behind. The camera shutter is at the very front of the instrument just behind the dark slide, but the instrument remains reasonably light tight during daytime engineering, for instance. An Instrument Electronics Box (IEB) is attached to the side of KOSMOS to control the various aperture wheels and focus mechanisms, and to read back environmental as well as positioning information. The IEB communicates with the KOSMOS computer rack via a single Ethernet cable.

The 2Kx4K e2v CCDs are very similar between the Kitt Peak and Cerro Tololo versions of the instruments, including near identical CCD gain and read noise. The only significant difference is the longer hold time of the larger COSMOS dewar. The Digital Head Electronics (DHE) box for each is bolted directly to the dewar. The DHE is another name for the Torrent controller, and Torrent is specifically a small footprint implementation of the Monsoon array controller architecture. The DHE is connected to the KOSMOS rack via a fiber optics cable. The DHE also controls the KOSMOS shutter.

The NOCS software talks to the IEB via a software component called the NOCS Instrument Control Software (NICS). The instrument can be controlled or monitored directly via the NICS GUI, but for the purposes of the DHS this is normally accomplished via observing scripts as discussed in section 5. However the motors have been positioned, the corresponding metadata should be reliably delivered to DHS via NOCS. These data are read before each exposure and again afterward. If there is any ambiguity between the two values, the DHS KTM (Keyword Translation Module, see section 4) generally pays attention only to the post-metadata. Values for some specific keywords such as the telescope pointing are copied from the pre-metadata to permit the telescope to be slewed during CCD readout.

The Monsoon software talks to the Torrent/DHE via the facilities of the Pixel Acquisition Node (PAN), and one usually speaks of the PAN computer versus the DHS computer. DHS speaks to the PAN, not directly to the DHE, and indeed DHS is linked against the Monsoon libraries. PAN delivers metadata to DHS in addition to the pixel data and NOCS and PAN are linked against DHS libraries for sending metadata and pixel data to the collectors.

For the case of KOSMOS, the Monsoon PAN performs significant reordering of raw pixels emerging from the CCD amplifiers (see figure 1). As with all standard CCDs, each amplifier implements a bucket brigade of horizontal per-pixel shifts toward the lower outside corners of the device. Each row is followed by an overscan of 50 pixels and then a downward parallel vertical shift of the next row above into the readout register. Thus the raw pixels emitted from the right hand side of the chip will be flipped in the X-axis relative to the left hand side. However, in the case of KOSMOS the PAN does three things: 1) the natural "engineering order" flip of the right hand side of the detector is reversed, 2) the two sides of the CCD are joined together seamlessly into a single image array, and 3) the left and then right side 50-pixel overscan regions are appended to the far right hand side of the image. (Whether the resulting "CCD detector order" of the image array is correct on the sky also depends on how the optics illuminated the chip and how the image display software presents them.)

In addition to the fundamental architectural dependencies of DHS on Monsoon and the NOCS, DHS relies on various third party software packages to a greater or lesser degree. These include:

**X11** – The DHS Real-time Display is implemented using the familiar IRAF ximtool display. The DHS Supervisor GUI is an IRAF widget-server program. Other windows are xterms or xgterms. Both at runtime and to build these clients, access to X11 is required. 32-bit X11 libraries are included in `/dhs/lib32`.

**VNC** – Virtual Network Computing (VNC) provides an OS-independent way to fire up client-server windowing over the network. For DHS this provides a remote window into the DHS computer from the observer's computer, which itself is typically remotely hosted from a Mac mini screen. All the DHS-related virtual windows (the Supervisor GUI, the Real-time Display, and various terminal windows) are gathered into a single actual window. This is very convenient but can also be confusing since the X11 preferences may come from one machine and the VNC preferences and authentication from another, perhaps under a different Unix account name.

For DHS instances supporting more complex multi-detector instruments, and thus multiple PAN computers and of shared memory caches, for instance, the VNC paradigm provides an X display server on a single machine that is used via X display settings to gather all DHS windows from all computers to this common display. Observers and other clients (e.g., telescope staff or programmers) can view this display with vncviewer from any machine on the net, allowing a remote view of DHS for debugging or a single window in the observer environment.

**PVM3** – DHS is a message-bus architecture and has been layered on PVM, the Parallel Virtual Machine[6], since the original Mosaic project. A copy of PVM3 is included under the DHS lib directory.

**Tcl** – The DHS Keyword Translation Module (KTM) is a Tcl (Tool Command Language) program and the DHS Data Capture Agent (DCA) is linked against the Tcl library.

**CFITSIO** – To support the alternate non-mbus mode in PXF, when not creating data using the DHS DCA. In normal operation the FITS files are creating via the MOSDCA DSIM (Distributed Shared Image) routines.

**IRAF** remains ubiquitous on NOAO mountaintops. This is particularly true for DHS-supported instruments that create FITS MEF files since observer support for displaying and processing these is accomplished using the IRAF MSCRED package. DHS is linked against the IRAF CDL library to support the Real-time Display.

Finally there is various software infrastructure such as compilers and unix tools, and the “Save the Bits” queues for archive data capture.

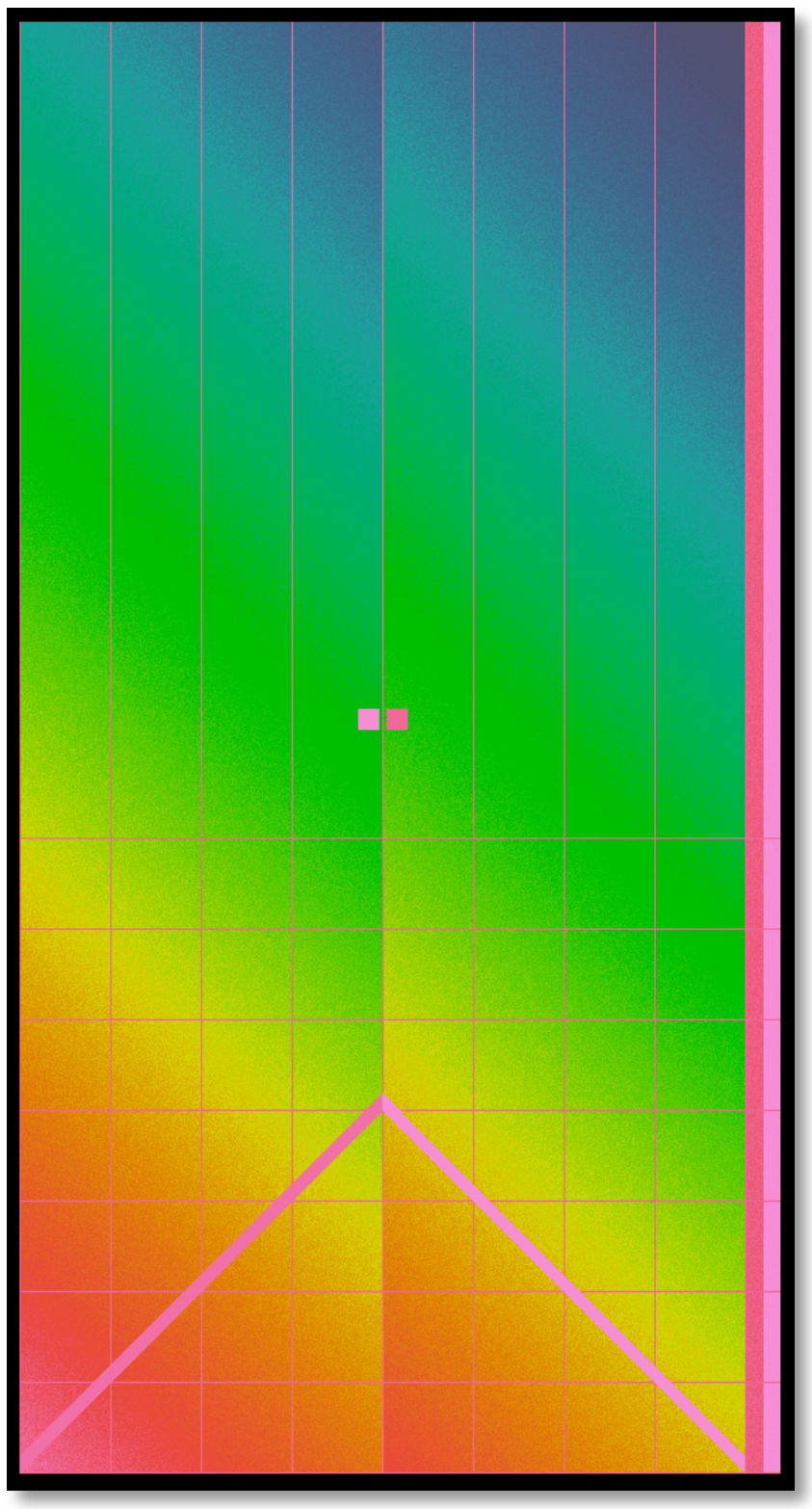

**Figure 1 – For KOSMOS using the e2v detector, the Monsoon PAN delivers a single array of pixels to the DHS collector. This array contains CCD pixels from both the left and right amplifiers followed by the left and then finally the right overscan. The right amplifier is flipped by the PAN such that the joined pixel array is correct on the CCD. (Amps are to the lower left and lower right.) The e2v CCD is 2048 pixels wide by 4096 pixels high and each overscan is 50 pixels, thus the full size is 2148x4096. KOSMOS pixels are 32-bit integers. The spectrograph dispersion runs vertically. The diagonal lines, squares near the center, and ruled grid are introduced in simulation mode.**

[6] http://www.netlib.org/pvm3/

## 3. DHS Structure

DHS is implemented in C-language components layered on a message-bus architecture and a central shared-memory cache. A supervisor orchestrates the actions of the other components and X11 GUIs provide the user interface. The ultimate objective of the DHS is to output individual exposures from the KOSMOS camera as Multi-Extension Format (MEF) FITS files, accessible to the astronomer and ingested into the NOAO Science Archive.[7]

The shaded area on the right hand side of figure 2 highlights the DHS command structure and data flow. DHS inputs are from NOCS and PAN, and DHS outputs are FITS files and to the user via a GUI and the Real-Time Display. Components surround the shared memory cache and exchange control messages via the common message bus; these components all exchange messages with the DHS Supervisor that then creates events acted upon by other components. There is generally no direct communication between the components (other than the data flow from PXF to MOSDCA).

On the other hand, the general pixel data flow is from PAN to Collector, through the SMC, taking a detour to the SMC Manager to be rectified, then to PXF which only then puts the data on the message bus. The MOSDCA assembles and outputs the final MEF FITS files including instrument and telescope metadata from the NOCS that passes through the Keyword Translation Module (KTM). While the Supervisor is orchestrating the events in the DHS via the message bus, the underlying data and metadata flow is via socket connections between PAN and Collector, for instance, and shared memory operations between the Collectors, SMC Manager and PXF.

A schematic of the KOSMOS spectrograph is indicated on the left hand side of figure 2 and the flow of events in the DHS follows from the observing cadence. The observer interacts with the NOCS software interfaces (scripts and GUIs) to construct and command exposure sequences. These may involve complex motions of the optical components of the spectrograph as well as telescope motions.

[7] http://portal-nvo.noao.edu

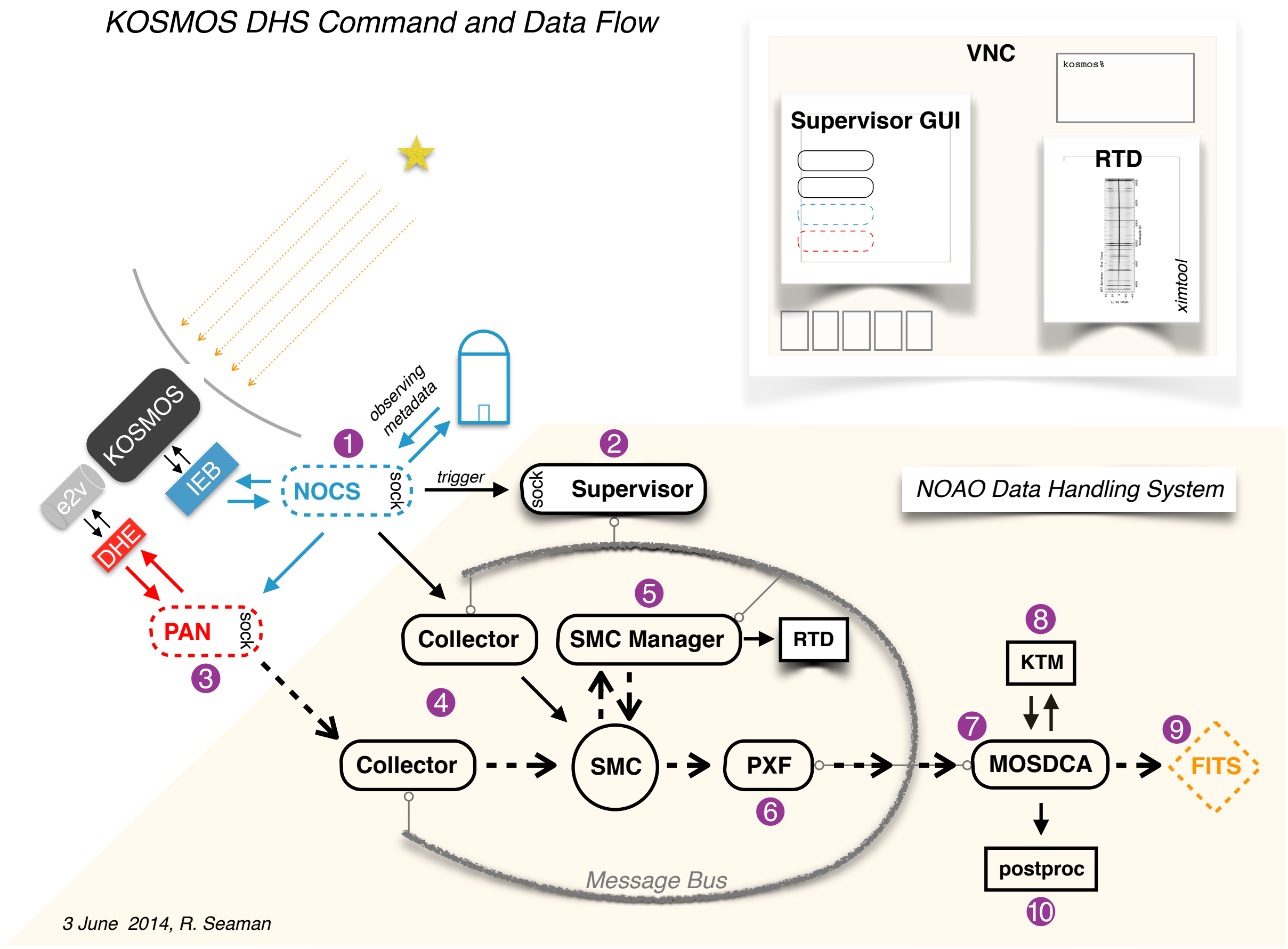

**Figure 2 – Command and data flow for the KOSMOS Data Handling System. DHS works in coordination with the NOAO Observation Control System (NOCS) and the Monsoon controller Pixel Acquisition Node(s) (PAN) to orchestrate KOSMOS observations and the CCD readout that follows. Note that the PAN and NOCS connect via sockets to the DHS Collectors and Supervisor, not via the Message Bus.**

The role of the DHS begins as each exposure ends:

1 The NOCS triggers DHS at end of the CCD readout.

2 The DHS Supervisor orchestrates collection of data and metadata from NOCS and PAN. Monitoring and control of DHS features is supported through the Supervisor GUI.

3 The PAN's control of DHE (and thus of the e2v CCD) has been coordinated with the NOCS control of instrument (and telescope, if appropriate).

4 The DHS Collectors store data and metadata from the PAN and NOCS in the Shared Memory Cache.

5 The SMC Manager rectifies the image, reading from and writing to the Shared Memory Cache, and then displays the rectified image in the Real-Time Display.

6 PXF loads the rectified data and associated metadata onto the Message Bus.

7 The Mosaic Data Capture Agent reads the data and metadata from the Message Bus. KOSMOS is a relatively simple chore for MOSDCA which is capable of combining data from multiple CCDs into large multi-extension FITS files.

8 The various sources of metadata are processed through the Keyword Translation Module, creating both the primary and extension FITS headers.

9 The new multi-extension FITS file is written to DHS computer's disk, one FITS Header-Data Unit (HDU) per CCD amplifier.

10 A postprocessing script is called with the pathname to new image. This script transfers the data to the observer’s computer disk, and triggers the initial archive transaction via the local “Save the Bits” queue. [Seaman 2005]

Figure 3 provides an alternate view of the NOCS / Monsoon / DHS interaction in the form of an activity diagram of the various messages exchanged. Time proceeds downward from the top, and the source and destination of each successive message are indicated by arrows between the various swim lanes. The left hand side of this diagram is quite complex, but the DHS interaction is fairly straightforward. DHS listens for data and metadata, and DHS receives a trigger at the end of the observation. As long as the pre and post exposure metadata, from both NOCS and Monsoon, and the pixel data from Monsoon are all tagged appropriately for each successive observation, the resulting “close exposure” event can asynchronously reassemble everything into a coherent data file.

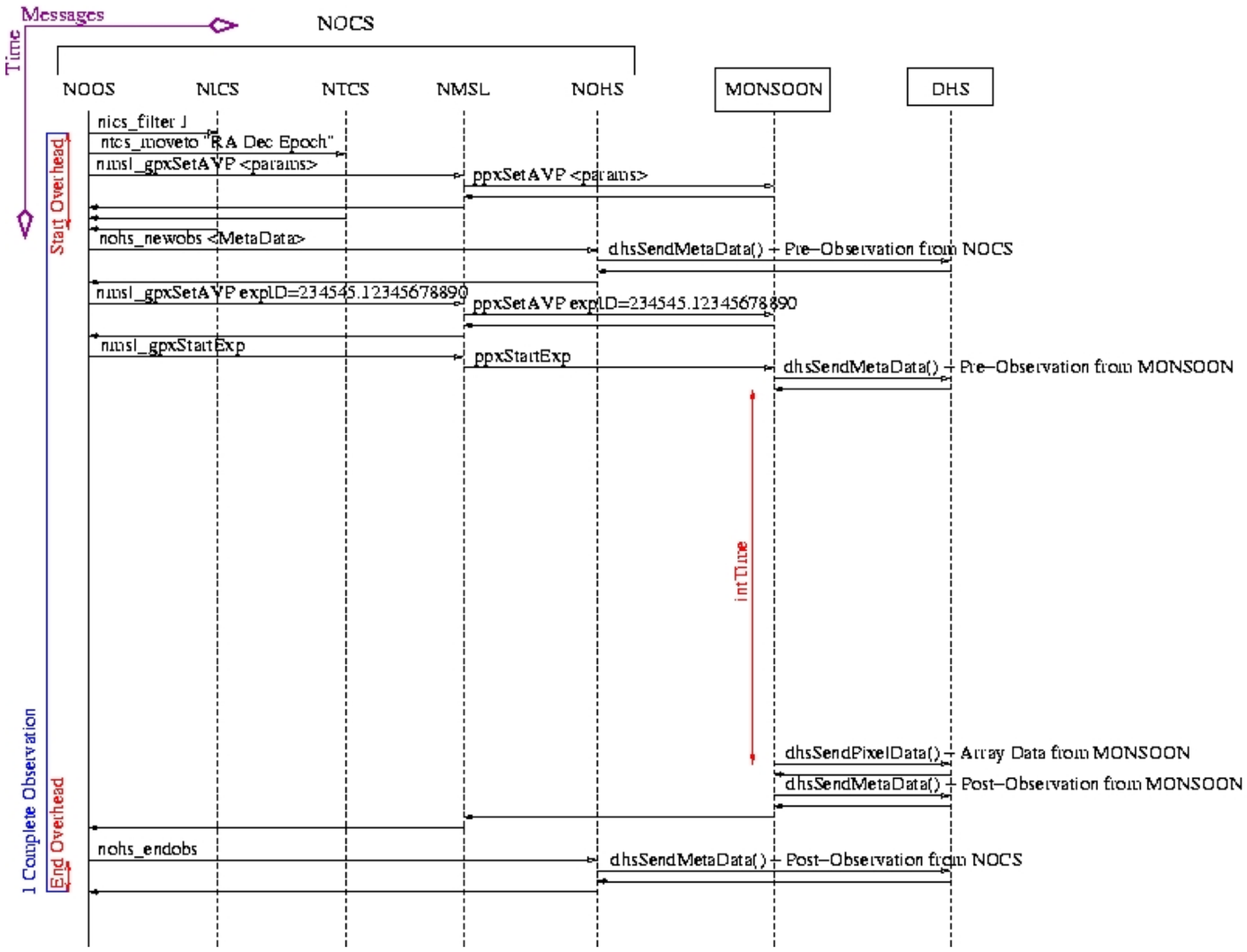

**Figure 3 – Message activity diagram for a typical KOSMOS exposure. [Daly 2010]**

In this typical exposure all commands that involve moving the telescope or adjusting the instrument occur at the beginning of the observation, but some complex observation types may well involve instrument motions throughout. One example is a multiple exposure when focusing the telescope.

## 4. DHS Software Components

The NOAO Data Handling System includes several components performing diverse functions. In addition to being monitored by the Supervisor GUI as discussed below, each of the components is also launched with text output redirected to a terminal window for logging of the nitty-gritty details. During normal observing these log windows are typically left iconified, but they can provide invaluable information for debugging purposes. The components can be started with command line access to various engineering functionality as described in the source code and implemented in the DHS startup script, `/dhs/bin/startdhs`.[8]

One peculiarity of DHS startup should be mentioned. During normal KOSMOS observing DHS is started by the NOCS, which has facilities for starting and stopping various components individually or together.[9] The NOCS is started on the main KOSMOS computer, `kosmos-4m` (or `cosmos-4m`), but the DHS runs on `kosmosdhs-4m` (or `cosmosdhs-4m`).[10] The way this is managed is in basically three steps: 1) the `/dhs` directory tree is synchronized (using rsync) from the main KOSMOS computer to the DHS computer (to pick up any recent changes), 2) a VNC instance is started locally, and 3) each DHS component is started remotely on the DHS computer. There are additional details in `startdhs`, but this remote execution is central. Note that if DHS or any of the other major NOCS or Monsoon components needs to be restarted for any reason, that the entire software system should be stopped (via "`nocs stop all`") and then restarted. This ensures that all the components are initialized properly and that various communications channels are established.

The major DHS components are:

**Supervisor** – The DHS Supervisor coordinates the activities of the other components by passing messages back and forth, by displaying status readouts to the user, and by providing an interface for user commands. Figure 4 shows the Supervisor GUI which monitors the various other component processes of the DHS and provides control access for those that need it, for instance the Real Time Display. Since the DHS is asynchronous, various activities can be going on for the exposure that is currently reading out as well as for a prior exposure being written to disk. The disk usage meter in the lower right hand corner is very useful. The Supervisor process executes on the main KOSMOS computer (along with ximtool used by the RTD), while all other DHS component processes run on the DHS computer.

---

[8] For the DHS simulator mode look at the script `/dhs/dev/zzsim`.

[9] Generally one begins a KOSMOS session with "`nocs start all`" and ends with "`nocs stop all`". A "`start all`" executes the `startdhs` script.

[10] The DHS Supervisor process runs on the main KOSMOS computer.

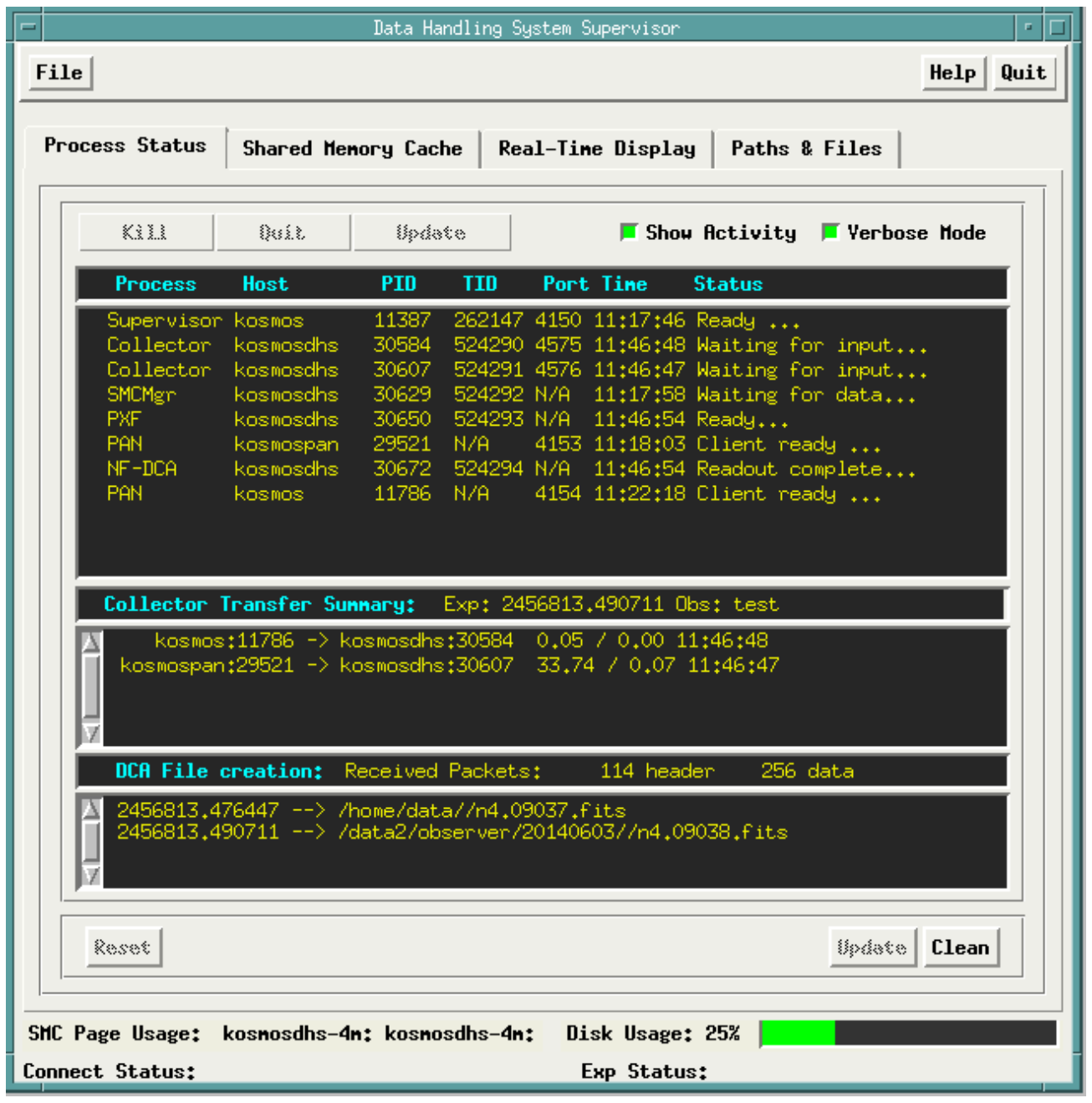

**Figure 4 – The NOAO Data Handling System Supervisor GUI.**

**Shared Memory Cache & SMC Manager** – The SMC Manager oversees the cached list of pages in shared memory that contain the data and metadata resulting for each exposure from the detector(s) and instrument. The SMC buffers these pages asynchronously to support cases of rapid readout. The SMC Manager is responsible for rectifying the pixel arrays in a instrument-dependent fashion (see `/dhs/dev/src/rectify.c`), passes images to the RTD, and provides garbage collection after images are finished processing through the DHS. The Shared Memory Cache tab in the Supervisor GUI allows monitoring the various SMC pages, which in normal operation should be retired as exposures are processed in turn.[11]

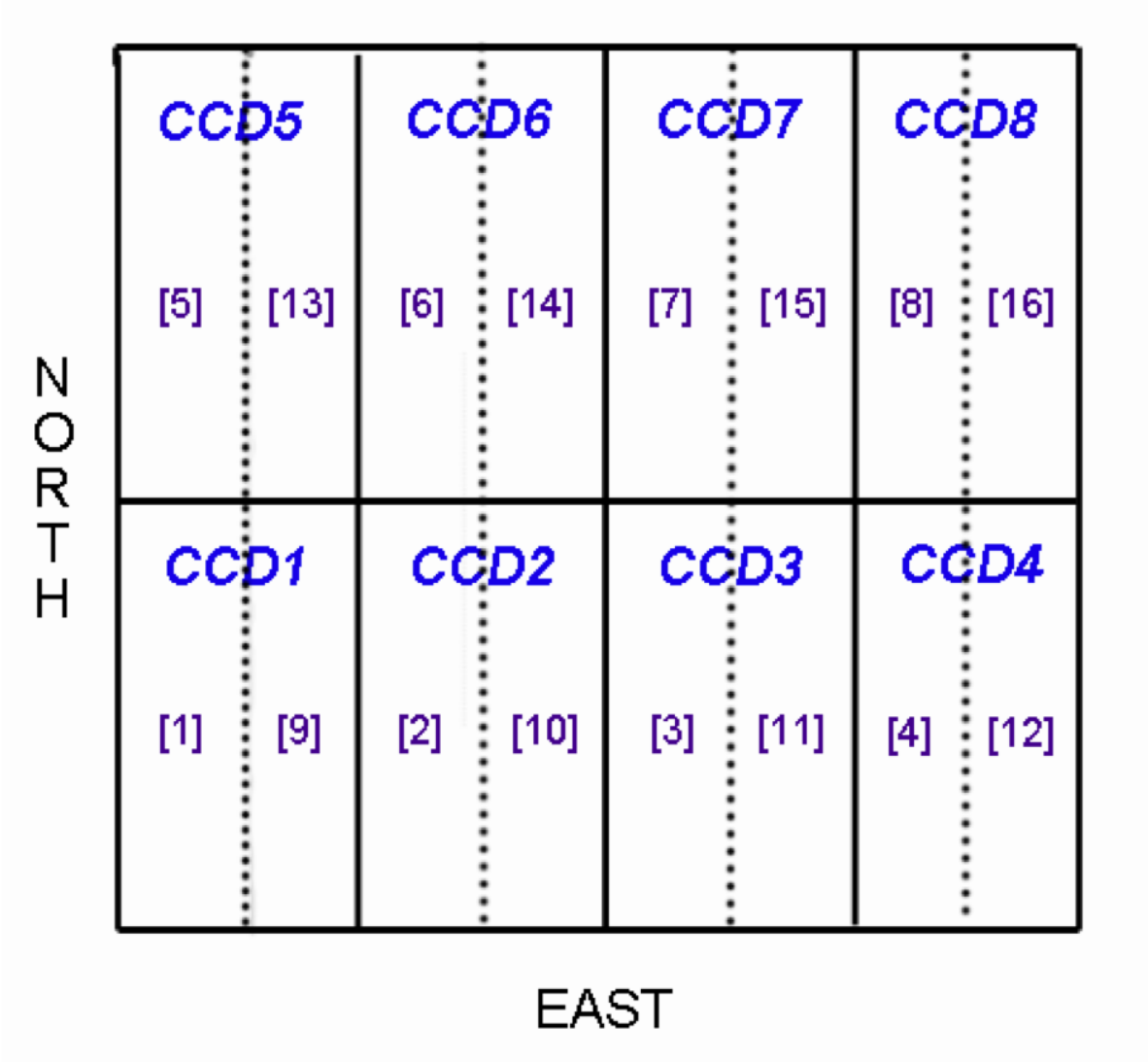

Figure 5 – The pixel rectification is more or less involved as required by the number of CCD or infrared detectors for a particular instrument and of the readout amplifier configuration of each. Mosaic-1.1 has 8 CCDs with 2 amps each, resulting in 16 pixel arrays. The top 8 are inverted in Y-orientation from the bottom 8, and the 8 right hand amps are flipped in X from the left hand 8. Flipping in both axes is equivalent to a 180 degree rotation. For detector purposes the KOSMOS e2v CCD is the same as CCD1 shown here.

---

[11] When first toggling to the SMC tab the instrument will be reported as Mosaic, not KOSMOS, since this is the default built into the GUI. This is normal and will update automatically as shared memory pages are allocated and released.

**Real Time Display** – Newly acquired images are displayed automatically as images are read out into memory. The observers may also interact with the RTD Ximtool using the IRAF display command, but often will run a separate instance of Ximtool or SAOImage DS-9 on the observer's computer. The Real Time Display tab in the Supervisor GUI provides a few controls for RTD behavior, for instance permitting trimming of the overscan to be toggled on and off.

**Collector** – The data that are handled by the DHS originate with one or more NOAO Monsoon array controllers that provide an electronic and software interface to optical and infrared detectors. A Torrent controller is a specific implementation of Monsoon. The Monsoon architecture passes data from the Torrent DHE (Digital Head Electronics) box via a fiber optic cable to a Pixel Acquisition Node (PAN), implemented in software on a Linux computer similar to that running the DHS. The job of the Collector is to receive data from the PAN and write it to the SMC. The data stream from the PAN includes pre-metadata, then pixel data, and lastly post-metadata. As shown in figure 3, one KOSMOS Collector instance also receives pre and post metadata from the NOCS. Each Collector provides a dedicated connection to either the NOCS or the PAN.[12]

**PXF** – PXF, which may sometimes be referred to as Picfeed, is the software component that reads the rectified pixel data from the SMC, along with associated metadata, and writes packets to the DHS message bus.

**The Collector, Shared Memory Cache, SMC Manager, and PXF must be deployed on the same computer host.** There may be multiple Collectors on the same DHS host as with KOSMOS, or there may be multiple complete sets of these four components on multiple hosts to distribute the load, but each such host must have all of these four components. As with the KOSMOS configuration the Supervisor can be on a separate host, and this can also apply to the Data Capture agent, described next.

**Mosaic Data Capture Agent** – The MOSDCA listens to the message bus and writes Multi-Extension FORMAT (MEF) FITS files to disk. For KOSMOS it need only handle one CCD and with the e2v detector only two amplifiers. The proposed LBNL detector has four amplifiers, and other instruments multiple chips. MOSDCA sorts out all that complexity. Metadata are passed to the Keyword Translation Module for formatting into header keywords [Seaman 2014].

**Keyword Translation Module** – The KTM is a Tcl-based instrument-specific component that implements a mapping of the Monsoon attribute-value pair (AVP) metadata to FITS header keywords. Tcl, the Tool Command Language, permits quite complex mapping of metadata as needed to translate CCD and instrument engineering information into science header keywords. For instance, the KTM provides a basic FITS WCS and computes diverse other header keyword values.

For KOSMOS and COSMOS, both instruments currently[13] use the same TCL code in two modules, `kosmos.tcl` (symlinked as `cosmos.tcl`), which references NOCS and PAN map files (text files relating metadata as known to those components) as well as the file `kosmos.dat` containing various default values. Then `kosmos.tcl` calls `kosmossrc.tcl` (symlinked as `cosmossrc.tcl`), which contains the logic of the KTM.

[12] For instruments other than KOSMOS there may be multiple PAN instances and computers.

[13] Both KOSMOS and COSMOS have been commissioned with a two-amplifier 2kx4k e2v CCD, but there are plans to also provide a four-amplifier LNBL detector for each. The KTM and other components will need to be updated at that point.

## 5. Observing with KOSMOS

Detailed instructions for observing with KOSMOS are provided elsewhere [Points 2014], including various DHS features accessible via the Supervisor GUI.[14] A brief overview is provided here to aid understanding of the DHS system concepts. While the two instruments, KOSMOS and COSMOS, are quite similar, details of the observatory control rooms and observing environments will differ and will not be described here. The KOSMOS software (meaning NOCS, Monsoon, DHS, and related packages) may be started via a desktop icon or from a command line prompt with the command "`nocs start all`" – consult the appropriate documentation for recommended usage at a particular site.

On whichever mountaintop, the KOSMOS observing console consists of several computer monitors providing feedback on this highly configurable instrument. As the numerous KOSMOS windows open during instrument startup (this may take a few minutes to complete), the DHS (figure 5) will appear and should be dragged to a convenient monitor location. Everything related to operating the DHS appears in the common VNC window.

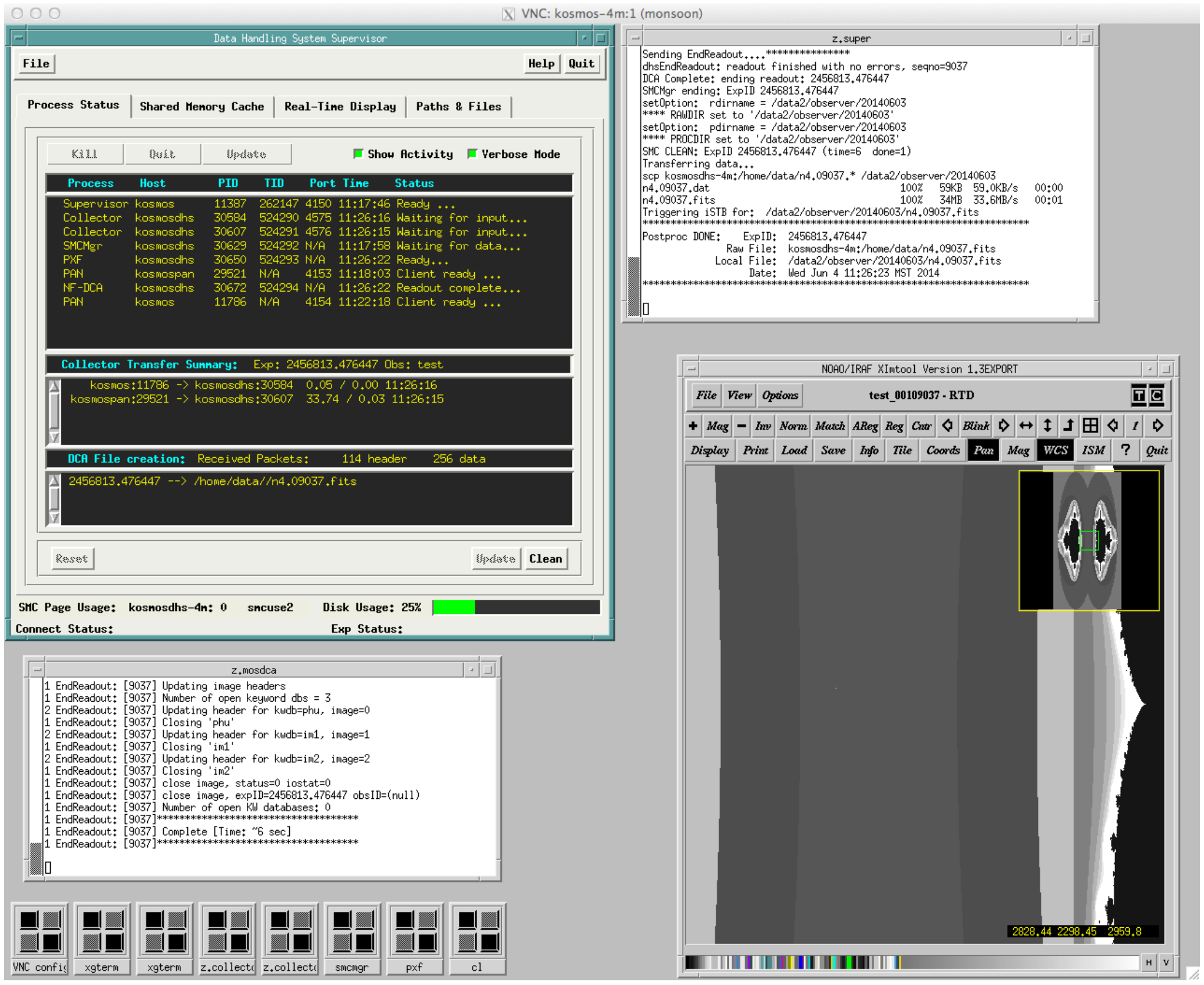

**Figure 6 – The DHS GUIs are collected into a single VNC window.**

[14] It can be helpful to compare KOSMOS with other Monsoon-based instruments that use the NOCS and DHS. The Mosaic-1.1 User Manual [Schweiker 2011] is particularly relevant.

The KOSMOS software packages are deployed on three separate computer hosts, with DHS executing on `kosmosdhs-4m` (or `cosmosdhs-4m`)[15] and the Monsoon PAN on `kosmospan-4m` (or `cosmospan-4m`). Only the single DHS VNC window generally talks to the DHS computer, and Monsoon generally operates without a user interface at all. All other windows will be connected to process running on the main KOSMOS computer, `kosmos-4m` (or `cosmos-4m`). These include quite a large number of NOCS-related windows, only a few of which will be discussed here. NGUI (figure 7) is a script editor for the KOSMOS observing scripts. This will typically be the first window that an observer interacts with to command a KOSMOS exposure. The DHS Real-Time Display will be the final window interaction for a given exposure.

After starting the KOSMOS software, NOCS and Monsoon as well as DHS, the first and perhaps only interaction the observer has with DHS Supervisor GUI may be to set the root name for the newly acquired images. This is done via the "Paths & Files" tab, see figure 6. Just backspace over the previous value and type a name pertinent to the night's observations.[16] The Apply button **must** be pressed to ensure that the new value takes effect. This new name should take effect for the next image that is processed by the DHS and should remain in effect through subsequent system restarts.

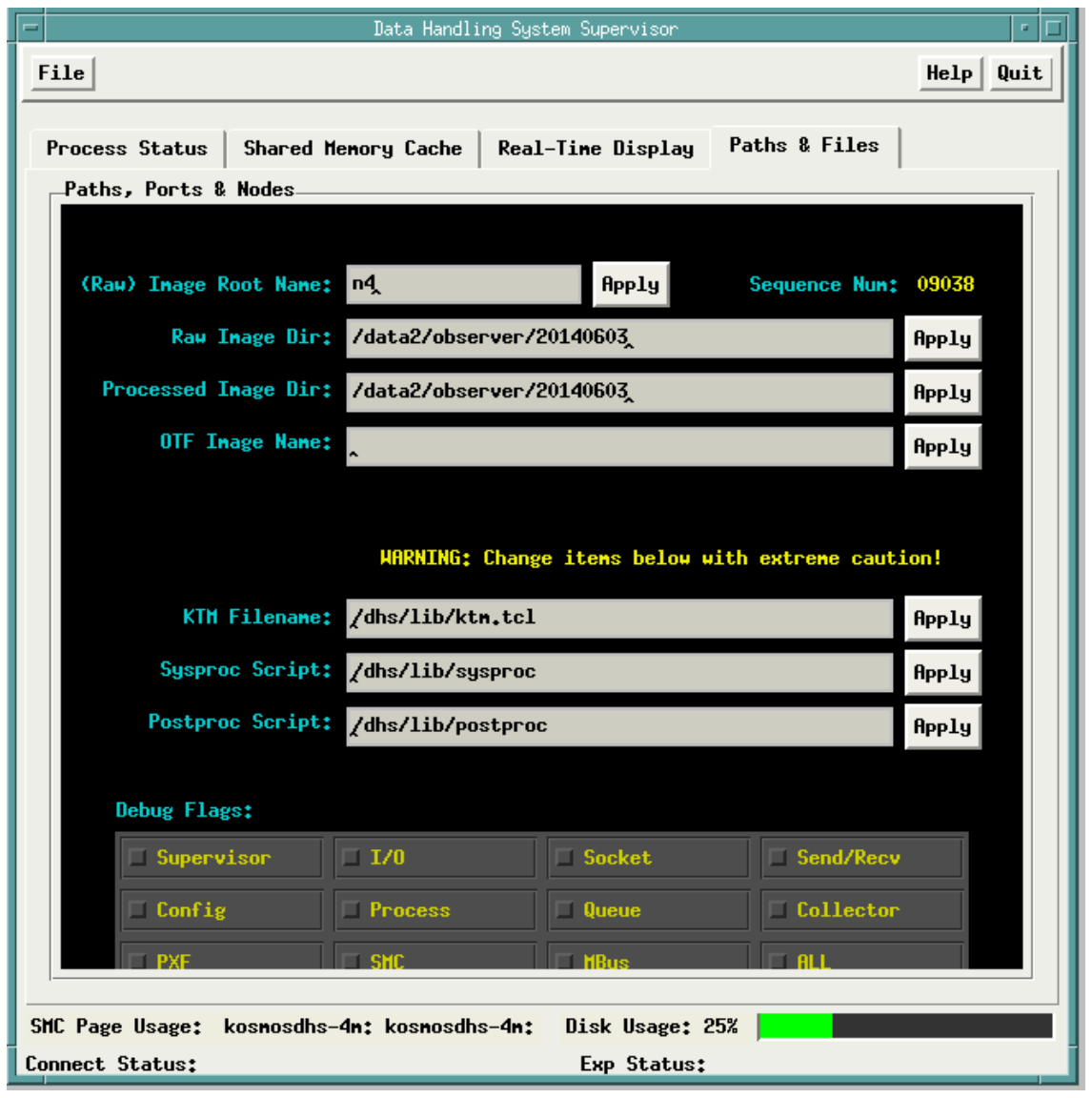

**Figure 6 – Tabs provide access to various DHS features, for instance, to change the root name for newly acquired images. Generally there are few knobs that an observer needs to twiddle.**

At this point KOSMOS should be ready to use.[17] The normal flow of activities is to create observing scripts using NGUI (see figure 7), and execute them from the unix command line (under the observer account on the main KOSMOS computer). These scripts can be created by the observer in advance, perhaps even at his or her home institution. If created on the mountaintop (or in the lab), they appear in the `exec` subdirectory of the observer's home directory and are executed like any other Unix command scripts.

---

[15] The DHS Supervisor executes on kosmos-4m (or cosmos-4m) along with VNC. All DHS processes run under the monsoon account on whichever host.

[16] The KOSMOS DHS will supply a period (".") in between the root name and the serial number that will be appended with each successive exposure.

[17] See the KOSMOS manual for issues of powering up the instrument and detector. We assume here that such prerequisites have been dealt with.

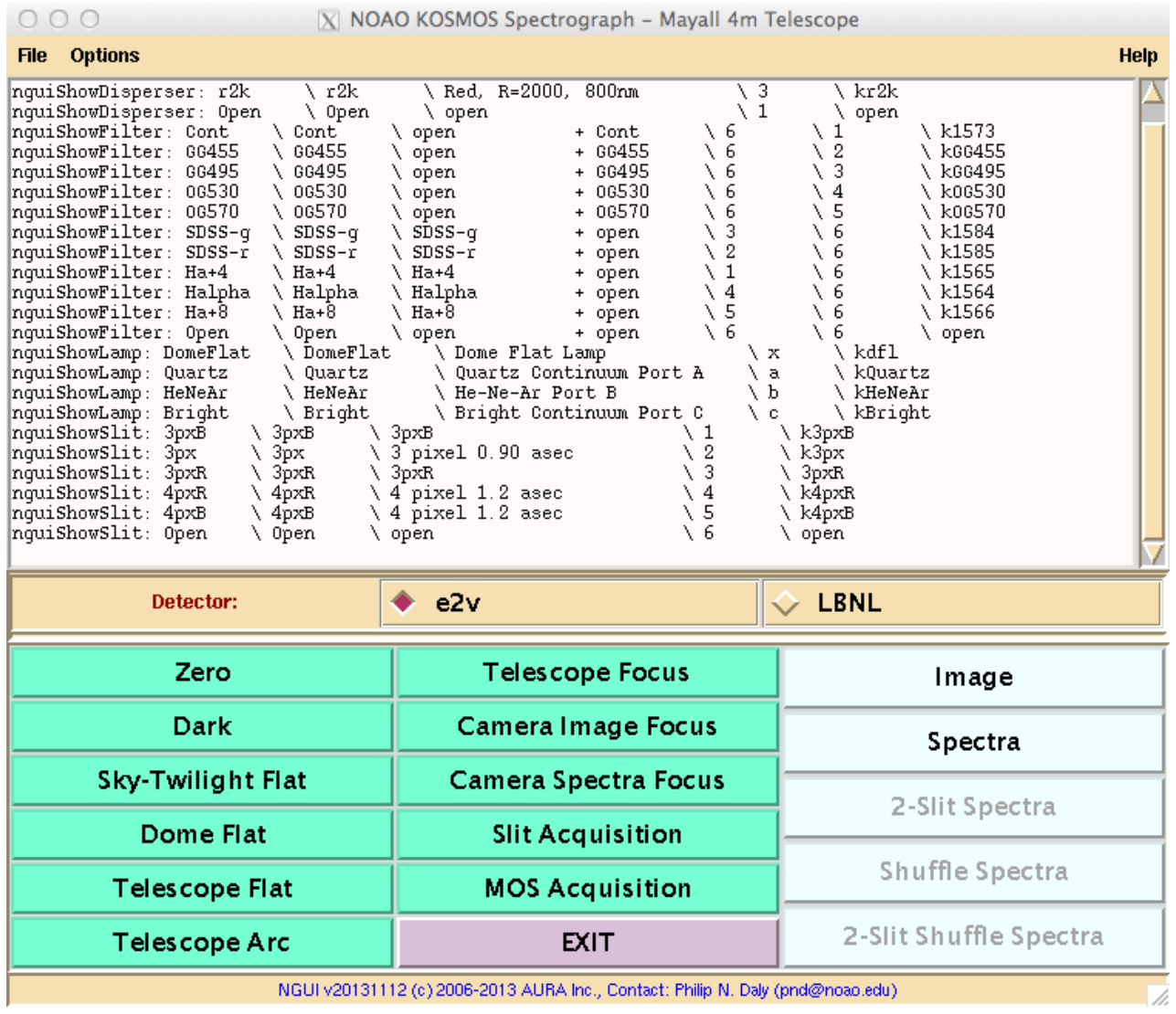

Figure 7 – NOCS provides a script editor, NGUI, for crafting often complex observation sequences. All KOSMOS exposures result from running one of these scripts, and it is the final CCD readout commanded by a script that triggers the interaction with DHS.

After selecting an observation type from the main NGUI window, a pop-up window (see figure 8) appears and allows setting values for the various parameters, exposure time, number of exposures, and so forth.

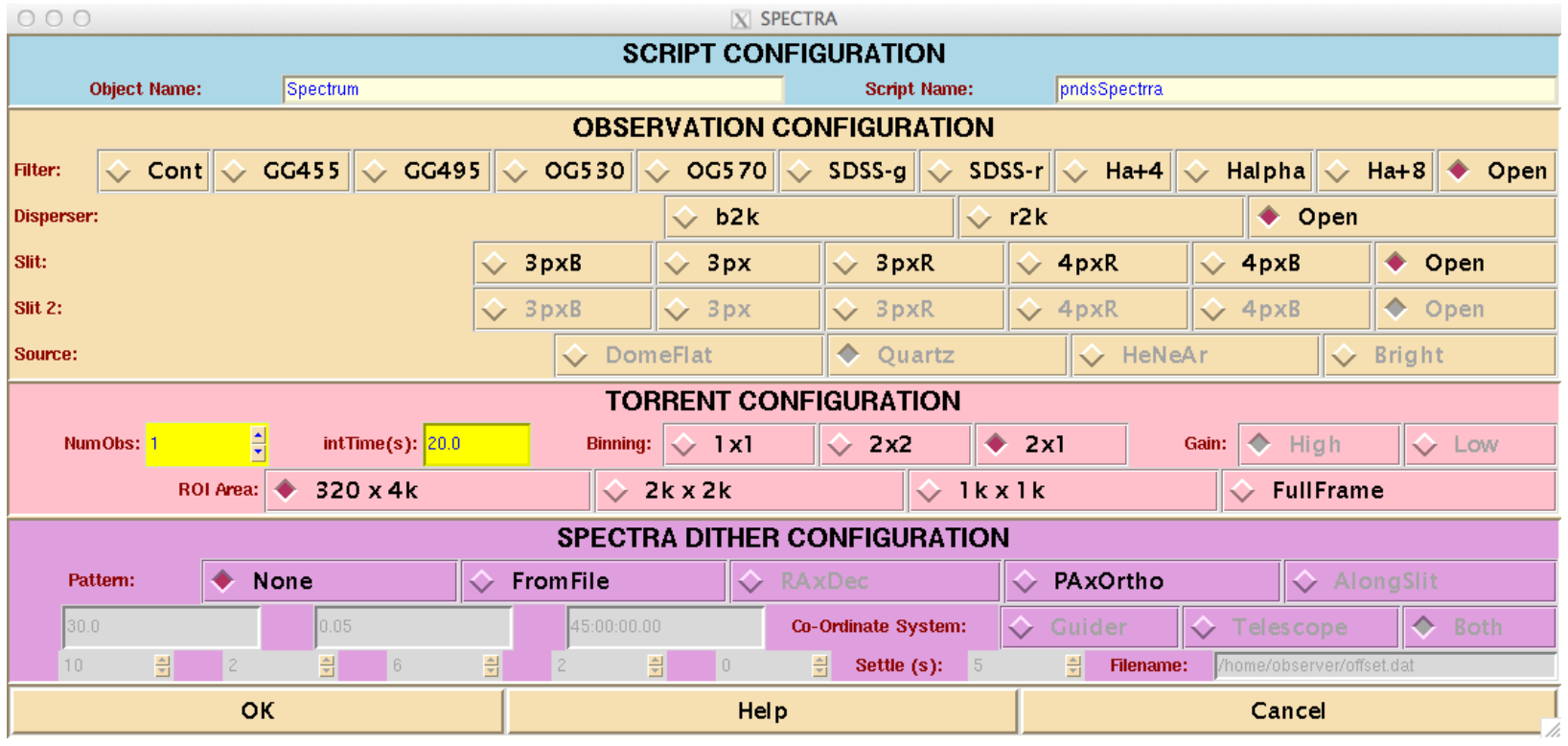

Figure 8 – Each exposure type has tailored script options. Some parameters such as the selection of the Region-of-Interest (ROI) or binning are common to all. Here a 320 x 4096 pixel ROI is selected with 2x1 binning (that is, binned by 2 spatially and unbinned in dispersion). These parameters are passed to the Torrent controller, which will later pass the metadata to DHS to permit correctly formatting the data arrays.

Some observation parameters are specific to a particular observation type, or for instance the filter is irrelevant to a zero (bias) exposure. Others apply to all types. In particular this is true of the Region-of-Interest (ROI) and CCD binning parameters. ROI is a new feature of DHS implemented for KOSMOS, as is the ability for different binning factors on each axis.

**Table 2 – Size in X/Y pixels of the *per-amplifier* CCD readout including overscan pixels for Region-of-Interest (ROI) and binning combinations. The full KOSMOS image is two 32-bit FITS extensions, each this size.**

| `NOCROIPT` | **1 by 1** | **2 by 1** | **2 by 2** |
|---|---|---|---|
| `FullFrame` | 1074 x 4096 | 562 x 4096 | 562 x 2048 |
| `2kx2k` | 1074 x 2048 | 562 x 2048 | 562 x 1024 |
| `320x4k` | 210 x 4096 | 130 x 4096 | 130 x 2048 |
| `1kx1k` | 562 x 1024 | 306 x 1024 | 306 x 512 |

The different ROI configurations supported by KOSMOS are listed in table 2 with the single amplifier image sizes resulting from each combination of an ROI and the X/Y binning factors. Note that "2 by 1" binning means unbinned in the dispersion direction and binned by 2 spatially along the slit. Two different pairs of configurations are highlighted that are not distinguishable merely from the size of the images. These have to be inferred from additional information in the DHS code (see `/dhs/collector/colPanHandler.c`), that is, from the X/Y pixel offsets as shown in table 3. If additional ROIs and/or additional binning combinations are implemented there are likely to be more of these ambiguous collisions in file size. It is left as an exercise for the interested student to identify one other that would appear given these ROIs, but with "1 by 2" binning.

Constraints placed on ROI geometry by the Torrent controller require that the ROI subsections for dual amplifiers must be mirrored relative to each other. For an example see the RTD in figure 11, which is a zoomed-in perspective on a 320x4k ROI, binned 2 by 1.[18] As a result all KOSMOS ROIs are regions centered on the e2V CCD detector. Table 3 describes the geometry of the four current ROI options, along with two additional proposed ROIs that are variations on the 1kx1k pattern but shifted vertically to aid in target acquisition in either the blue or red end of the spectrum. These proposed ROIs remain centered along the X-axis. Note the potential confusion from the terminology of rows and columns versus X and Y.

**Table 3 – Each ROI pattern corresponds to a unique combination of X/Y size (per-amplifier and without overscan) and X/Y offset in pixels from the lower left hand corner. The row/column nomenclature is reversed from the usual X/Y coordinates. Also included here are the proposed variations of the 1kx1k ROI shifted along the dispersion axis toward the red and blue.**

| `NOCROIPT` | `NOCROIRZ` | `NOCROICZ` | `NOCROIRS` | `NOCROICS` |
|---|---|---|---|---|
| `FullFrame` | 4096 | 1024 | 0 | 0 |
| `2kx2k` | 2048 | 1024 | 1024 | 0 |
| `320x4k` | 4096 | 160 | 0 | 864 |
| `1kx1k` *center* | 1024 | 512 | 1536 | 512 |
| `R1kx1k` *red* | 1024 | 512 | 2048 | 512 |
| `B1kx1k` *blue* | 1024 | 512 | 1024 | 512 |

After creating a KOSMOS observing script with the desired parameters, and executing either a single exposure or a multiple exposure sequence, the Monsoon software commands the DHE to take one or more CCD exposures. DHS at this point is waiting for data from the PAN and for metadata from both the PAN and NOCS (and ultimately from the KOSMOS instrument via the

[18] The grid and the small rectangles are simulated structure in the illuminated pixels, not the mirrored geometry.

IEB and from the telescope control system and other observatory systems). The "NOCS Monsoon Supervisor Layer" (NMSL) window provides feedback during the exposure and CCD readout (see figure 9). Two things to be aware of here[19] are that the exposure timer, in particular the timing of the readout, is only approximately correct. Rarely will it count down to zero, and one will soon get used to what values to expect for typical configurations, such as ROIs.

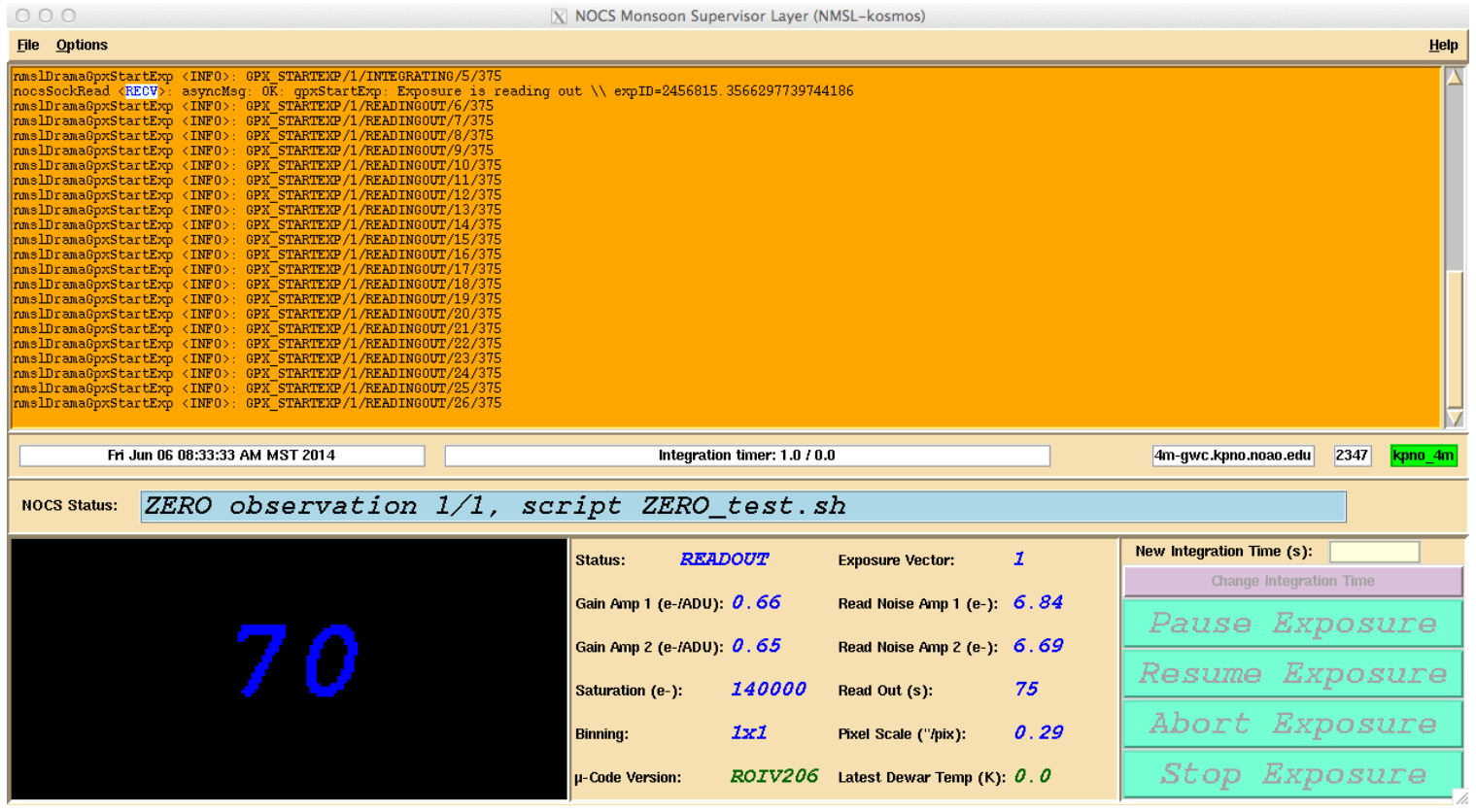

**Figure 9 – DHS works in concert with the various NOCS windows. The complete KOSMOS NOCS environment is quite complex and includes several other GUI interfaces for either monitoring or controlling the instrument or telescope. The orange color here indicates that the Monsoon PAN is in simulation mode.**

The other thing is a more fundamental question, namely: "Am I actually talking to the detector?" The orange color of the NMSL window is the tell-tale indication that the PAN is in simulation mode and that no actual CCD data are flowing from the DHE to the DHS. This can happen in at least a couple of ways. First, is the DHE actually plugged in and turned on? Second, is the fiber optic connection between the DHE and the PAN computer working correctly? In both cases the KOSMOS software must be halted (`nocs stop all`)[20] and the issue must be cleared. Consult the KOSMOS Instrument Manual for details on power-cycling the DHE.

If the power proves to be on, the issue is likely the fiber link and the fix is "`~/bin/fixlink`" (from the observer account on the main KOSMOS computer). Fixlink also nukes just about everything and can be regarded as the command to try if all else fails.

When the CCD readout finishes a picture should immediately appear in the RTD (in the lower left hand corner of figure 5). DHS is asynchronous such that even though a picture has been displayed the corresponding FITS file may not have been written to disk. The image in the RTD is provided by the SMC Manager directly from the shared memory cache. For KOSMOS, however, the FITS file should follow only a few seconds after the readout. The e2v CCD used with KOSMOS generates 32-bit integer pixels via an 18-bit A/D. Thus a 2148x4096 full frame image will be about 35 megabytes uncompressed.

The image displayed in the RTD is composited from the rectified image in shared memory. The "Real-Time Display" tab in the Supervisor GUI provides a few configuration options, for instance whether to trim the bias overscan pixels from the display. Since an 8-bit ximtool is used by the RTD the color mapping will change as that window receives and loses cursor focus. To interact with the full capabilities of the display, for instance to use the IRAF imexamine task, the image should be redisplayed from the FITS file. This will provide accurate pixel values (the RTD uses a z-scaled approximation) as well as access to the FITS WCS (world coordinate systems). Since

[19] A third issue sometimes occurs when starting NMSL. If errors are displayed in the NMSL window right after starting the KOSMOS software, try an "nmslreset" (as observer on the main KOSMOS computer).
[20] A "nocs stop hardware" (and later a "nocs start hardware") may also be needed.

DHS creates multi-extension format (MEF) FITS, the correct command is mscdisplay from the IRAF mscred package. A comparison of displaying a (simulated) image using mscdisplay versus separately displaying the two image extensions using IRAF "display" is in figure 10.

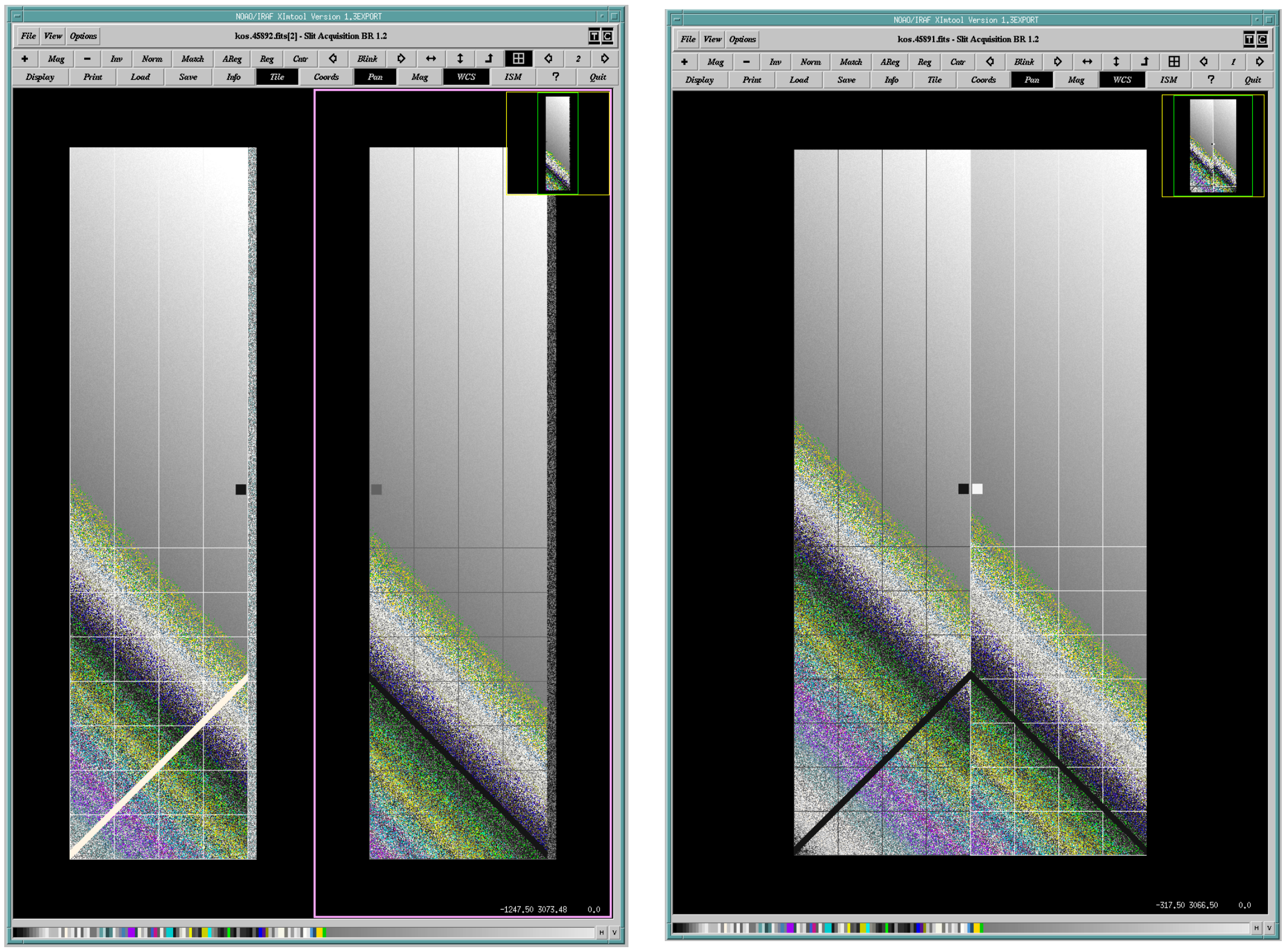

**Figure 10 – Comparison of displaying the individual KOSMOS FITS image extensions (HDUs) on the left versus mscdisplaying both on the right (simulated data). The correct order of the pixels on the CCD is preserved in both cases as delivered by the PAN, not the readout order from the amplifier. (The colorful noise is simply an artifact of the 8-bit display and will revert to grayscale when the cursor is in the window.)**

The final act of the DHS Supervisor for this exposure is to call `/dhs/lib/postproc`.[21] Remember that the Supervisor executes on the main KOSMOS computer, not on the DHS computer. Postproc has two main responsibilities: 1) arrange to copy the newly acquired image back from the DHS computer to the main KOSMOS observing computer, and 2) send the image to the NOAO Science Archive using the iSTB queue configured on the DHS host. In the lab this latter step will be omitted. The DHS also supports a separate `sysproc` script as shown in figure 6, however this is not currently used for KOSMOS other than for logging. The idea behind separating the two scripts is to permit more critical tasks to be placed in `sysproc` and perhaps permit each observer to modify postproc to suit his or her needs.

[21] While the postproc script is user configurable via the "Paths & Files" tab of the Supervisor, changing the default is a very poor idea.

## 6. Observation Simulator

The previous section described data-taking with KOSMOS. That mode requires access to the KOSMOS instrument and IEB, to an e2v CCD detector and the attached Torrent DHE, and to the KOSMOS computer rack.[22] At the very least one of the KOSMOS / COSMOS computer racks is required to run in PAN simulator mode. There are complex interactions between the various NOCS, Monsoon, and DHS components running on three different computers under two different Unix accounts, and there is no substitute for the full system to exercise and verify functionality approximating normal observing conditions.

On the other hand, if the goal is to implement or provide maintenance for the Data Handling System on its own, negotiating access to the mountaintop KOSMOS resources may be unnecessary and burdensome. To support these cases DHS has its own observing simulator as shown in figure 11.

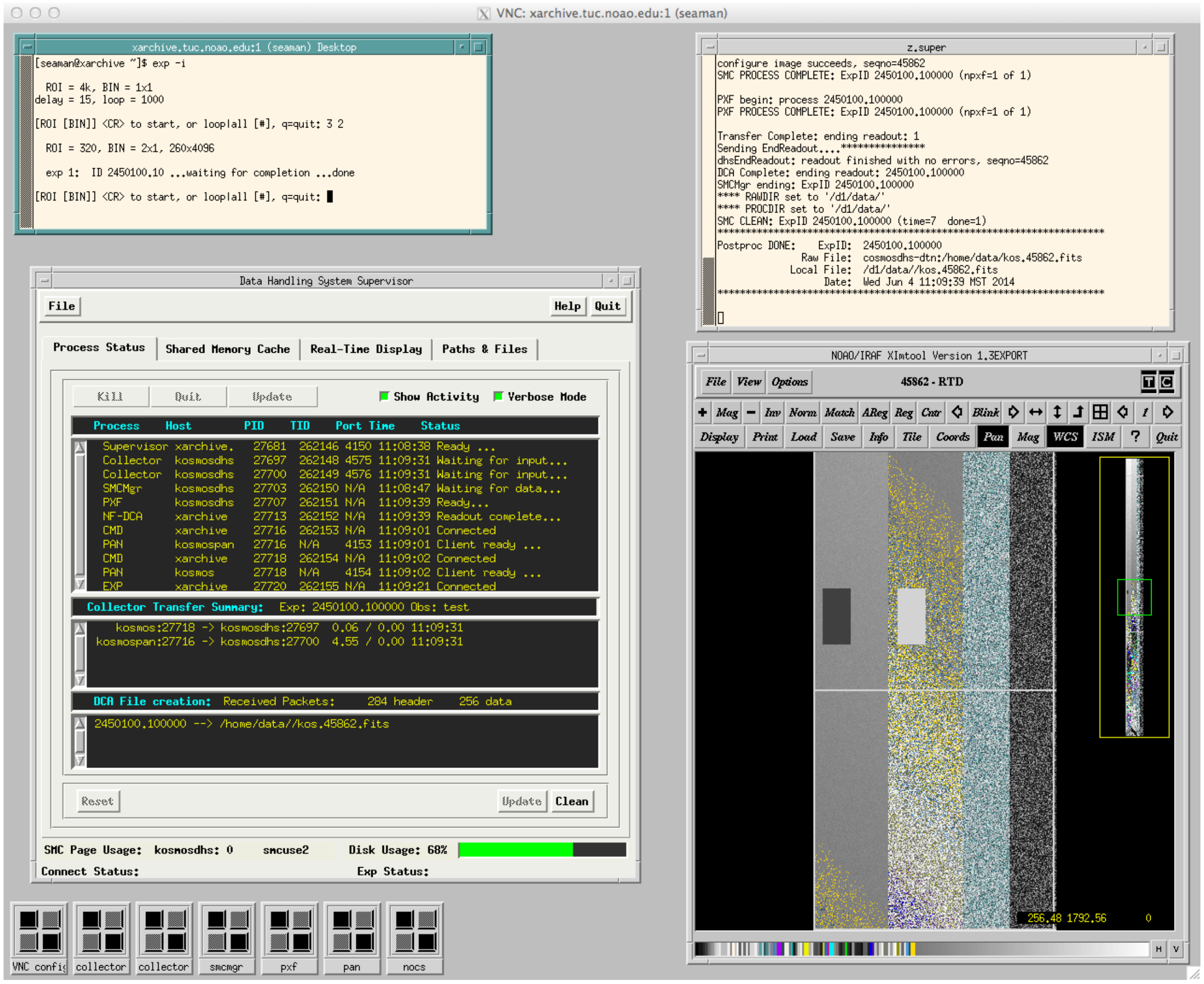

**Figure 11 – When operated in simulator mode, DHS runs standalone.**

[22] For full functionality it would also need to be nighttime at one of the telescopes.

Referring back to figure 2 on page 5, one sees that DHS requires inputs from the NOCS and the PAN, and writes its FITS output to local disk (local to the DHS computer, that is). Simulating an exposure is therefore an exercise in providing the right inputs, not in modifying the DHS components themselves.

Three processes work together to provide this input. These are simplified substitutes for the NOCS and Monsoon PAN, in addition to an EXP command that stands-in for an NGUI script. The source code for these is `nocs.c`, `pan.c`, and `exp.c` in `/dhs/test`. The simplest way to see what they do is to look at figure 3 on page 7. The pseudo-NOCS and PAN interactions implement the pre, post, and pixel messages sent to the DHS, and EXP triggers each exposure or exposure sequence. These three source files are a good place to start to understand the nature of the transactions between the NOCS and PAN and the DHS. The files `nocsHdr.h` and `panHdr.h` define a default set of AVP metadata (attribute value pairs) that will be parsed by the standard KTM.

To run the DHS simulator, follow the instructions in section 7 to check out and build a copy of DHS. Then run the simulator script from a suitable instance of X11 running under a 32-bit Linux OS, e.g.:

```
% ssh –X –Y xarchive

xarchive% /dhs/test/zzsim
```

*<lots of messages>*

The DHS VNC should appear, resembling figure 11. In the xterm in the upper left hand corner of the VNC, type:

```
% /dhs/test/exp -i
```

This starts the exposure simulator in an interactive mode. The prompt describes the available commands:

```
[ROI [BIN]] <CR> to start, or loop|all [#], q=quit:
```

Which is to say that the any of the supported KOSMOS ROIs can be selected:

```
ROI must be one of 4k, 2k, 1k or 320[23]
BIN must be one of 1x1, 2x1, 2x2
```

Pressing enter (`<CR>`) repeats the last exposure, typing “loop 10” does a sequence of 10, and the command “all” cycles through all combinations of ROI and binning, several times if provided with an argument. Type “q” to quit the exposure simulator. Dismiss the VNC (and perhaps kill zzsim in the original window) to exit DHS.

Generally this interaction will be useful for implementing new features in DHS, for instance by adding simulator support for the proposed R1kx1k and B1kx1k ROIs, and verifying this against the necessary changes to the handling of these cases in the core DHS components.

---

[23] Keystroke shortcuts are implemented: 1 through 4 for ROI, and 1, 2 or 4 (that is, 2x2) for binning.

It should be noted, however, that what the DHS simulator does not simulate are the interactions of the KOSMOS instrument operations at the telescope. For instance, in simulator mode all of the processes are executed locally such that the substitute NOCS and PAN commands run on the same machine as the DHS components, as do the DHS Supervisor, RTD, and VNC that normally run on the main KOSMOS computer, not the DHS host. If the issue being investigated is something to do with the KOSMOS rack or network, only the mountain environment will do. Similarly the entire startdhs procedure is replaced with a different tailored script. And, of course, issues of moving aperture wheels in the instrument or of telescope metadata, etc, will require something approximating the full KOSMOS system.

## 7. Installation and Troubleshooting

The KOSMOS DHS software is available as a git[24] repository on bitbucket.[25] The basic installation involves just a few steps. Since DHS is designed to operate a complex scientific instrument in collaboration with Monsoon and the NOCS software, configuring the software requires understanding the complete observatory context. The first step is to identify an appropriate computer host with a 32-bit Linux operating system:

```
% uname –m
i686
```

At NOAO the DHS is installed and runs under the monsoon account. The Monsoon libraries should be installed on the machine:

```
monsoon% cd
monsoon% pwd
/home/monsoon
monsoon% ls –ld Monsoon*
lrwxrwxrwx  1 monsoon users   28 Jun 26  2013 MonsoonInc ->
/home/monsoon/MonsoonSrc/inc
drwxrwxr-x 12 monsoon users 4096 Nov 27  2013 MonsoonSrc
```

Create an appropriate directory to contain the DHS software.

```
monsoon% pwd
/home/dhs
```

Clone the git repository into a new subdirectory:

```
monsoon% cd /home/dhs
monsoon% git clone git@bitbucket.org:rseaman/kosmos_dhs.git
```

Symlink the `kosmos_dhs` subdirectory to /dhs:

```
monsoon% su
root% ln –s /home/dhs/kosmos_dhs /dhs
```

---

[24] http://git-scm.com
[25] https://bitbucket.org

Make DHS:

```
monsoon% cd /dhs
monsoon% make
```

*(lots of messages)*

```
Update Done.
```

Add /dhs/bin to the monsoon account unix path. Rehash. Sections 5 and 6 provide instructions for running the software in the full KOSMOS environment (with Monsoon and NOCS) and in the DHS simulator environment, respectively.

Various troubleshooting comments have been made above, but are also collected here in more detail.

- On occasion the KOSMOS rack may require rebooting. The command for this is:

  ```
  observer% nocs reboot rack
  ```

  and should typically be issued under the observer account (like all `nocs` commands) on the main KOSMOS computer, `kosmos-4m` (or `cosmos-4m` or `–dtn` versions). It is hard to give precise guidance for when a reboot may be necessary, but this most likely would occur when actively working on the software and somehow the run-time context has gotten very confused. A reboot takes several minutes to finish and care should be taken that all three KOSMOS hosts (DHS and PAN as well as the main computer) have come back up again. Rebooting the rack is about the last thing to try.

- More frequently a run-time snafu will arise that can be cleared using the `fixlink` command. This combines all of the incantations that restore run-time context for the major subsystems (Monsoon, NOCS, and DHS) to their defaults. It should only be issued when the KOSMOS software has been halted:

  ```
  observer% nocs stop all
  observer% nocs stop hardware

  observer% ~/bin/fixlink

  observer% nocs start hardware[26]
  observer% nocs start all
  ```

  Issuing a `fixlink` command is a good idea whenever the DHE is power-cycled.

- One of the things fixlink does is to issue the DHS `zzkpvm` command that restores a known context for the PVM software on which the message bus is layered:

  ```
  ssh monsoon@kosmos /dhs/bin/zzkpvm
  ```

[26] KOSMOS may be configured to start and stop the instrument hardware initialization separately from the Monsoon and DHS related initialization; consult the observing manual.

This is also executed by the startdhs script. With KOSMOS there are generally few instances to issue zzkpvm directly, but this might arise, for instance, if a programmer is actively working on issues related to the message bus.

- On occasion the NMSL window will start up in PAN simulator mode. This is indicated by the orange color of the window, and may be caused by the DHE being powered off or by the fiber link containing stale data. The cure is to follow the full fix link recipe above by stopping "all" and "hardware", running fixlink, and starting up from scratch.

- If NMSL is NOT orange, but if on starting up there are error messages (typically involving voltages) displayed in the window, typing `nmslreset` at the Unix prompt in the window from which the `nocs start all` command was issued will usually clear the problem. If such error messages persist, talk to the instrument engineers.

- Even during normal KOSMOS operations the countdown timer in the NMSL window is largely just suggestive of the actual timings involved. A particular exposure type, ROI and binning combination, however, should provide a very repeatable readout reading.

- Many observers will choose to run an instance of ximtool or saoimage ds9 separate from the RTD ximtool in the DHS window. In that case the image displayed will naturally have started with a copy of a KOSMOS FITS file and should provide accurate pixel values and coordinates using general IRAF or non-IRAF commands. If, however, an imexamine or other IRAF cursor readback command is issued on the DHS RTD, users should be aware that they may need to redisplay the image from the FITS file.

- During normal KOSMOS observing, when NOCS is started it will create a symbolic link, "`data`", to the night's data directory:

```
[observer@kosmos-4m ~]$ ls -l ~/data
… /home/observer/data -> /data2/observer/20140608
```

  If the KOSMOS software is not restarted at least daily, this symlink may not be updated, but the data should still be copied to the correct directory on the main KOSMOS computer. However, if you edit paths[27] in the Supervisor "`Paths & Files`" tab you **must** press the "`apply`" button.

- If the ximtool window in the DHS goes away (I've never seen this), short of a "`nocs stop all`" and "`nocs start all`", try executing `/dhs/x11iraf/ximtool/ximtool` at a monsoon user prompt on the main KOSMOS computer.

- If the DHS Supervisor window suddenly goes red, this indicates some error condition. If during software development it is likely pilot error and a full restart will be needed after correcting the coding issue. If during normal observing it likely means a communication issue with the PAN. KOSMOS pays close attention to the ROI and binning metadata[28] and if these don't match the actual data arrays that are received, the data handling context can become erroneous. In these cases it is likely that the SMS Manager has crashed and there will be error messages to that effect, and/or the smcmgr log window will vanish. In those cases a full restart is required.

[27] Generally not advisable.
[28] KOSMOS pays closer attention than either NEWFIRM or Mosaic-1.1, which do not support ROI readout or non-square pixel binning. Indeed, the 2kx4k chip size is hardwired for Mosaic-1.1 DHS.

- If the SMC Manager is still running, a recovery can be attempted.[29] First try clicking on the "Update Status" button on the DHS Supervisor "Shared Memory Cache" Page. Do messages update in the z.super window? Perhaps press the button again and see if the SMC entries vanish (are processed).

- At the observer account command prompt (meaning any NOCS script being executed needs to finish or be halted), type:

```
ditscmd nohs nohs_endobs
```

  then perform the "Update Status" recipe.

[29] These procedures are by analogy with NEWFIRM. After many thousands of commissioning images, such instances remain rare enough with KOSMOS that sufficient errors have not accumulated to characterize the best user responses in each case. A full NOCS stop and restart, perhaps power cycling the DHE and IEB, should restore proper operations.

## 8. References

1. Daly, P., “KOSMOS ICD 6.1 Instrument Controller to Data Handling System”, NOAO (2010).
2. Martini, P., et. al., “KOSMOS and COSMOS: New Facility Instruments for the NOAO 4-meter Telescopes”, to appear in *Ground-based and Airborne Instrumentation for Astronomy V*, Proc. SPIE 9147 (2014).
3. Points, S., Elias, J., Martini, P., & Beers, T., “C/KOSMOS Instrument Manual”, NOAO (2014).
4. Schweiker, H., et. al., “KPNO Mosaic-1.1 Imager User Manual”, NOAO (2011).
5. Seaman, R., Barg, I., & Zárate, N., “The NOAO Data Cache Initiative – Building a Distributed Online Datastore”, ASP Conf. Series v. 247, 679 (2005).
6. Seaman, R., KOSMOS Keyword Dictionary, NOAO, (2014).
7. Zárate, N. & Fitzpatrick, M., “The NOAO NEWFIRM Data Handling System”, ASP Conf. Series v. 394, 677 (2008).